\documentclass[11pt]{article}
\usepackage{amssymb,amsmath,fullpage,graphicx}
\def\[{\begin{equation}}
\def\]{\end{equation}}

\let\^=\hat
\let\~=\tilde

\makeatletter \numberwithin{equation}{section}

\begin{document}
\title{
Can parity-time-symmetric potentials support continuous families of
non-parity-time-symmetric solitons?}

\author{Jianke Yang \\
Department of Mathematics and Statistics \\ University of
Vermont,\\Burlington, VT 05401, USA }

\date{ }
\maketitle

\begin{abstract}

For the one-dimensional nonlinear Schr\"odinger equations with
parity-time (PT) symmetric potentials, it is shown that when a real
symmetric potential is perturbed by weak PT-symmetric perturbations,
continuous families of asymmetric solitary waves in the real
potential are destroyed. It is also shown that in the same model
with a general PT-symmetric potential, symmetry breaking of
PT-symmetric solitary waves do not occur. Based on these findings,
it is conjectured that one-dimensional PT-symmetric potentials
cannot support continuous families of non-PT-symmetric solitary
waves.

\end{abstract}

\section{Introduction}

Solitary waves play an important role in the dynamics of nonlinear
wave equations (Kivshar \& Agrawal 2003, Yang 2010). In conservative
systems, solitary waves generally exist as continuous families.
Familiar examples include the nonlinear Schr\"odinger equation with
or without external real potentials. In dissipative systems,
solitary waves generally exist as isolated solutions, with the
Ginzburg-–Landau equation as one of the best known examples
(Akhmediev \& Ankiewicz 2005). However, a recent discovery is that,
in dissipative but parity-time (PT) symmetric systems (Bender \&
Boettcher 1998), solitary waves can still exist as continuous
families, parameterized by their propagation constants. This exact
balance of continually deformed wave profiles in the presence of
gain and loss is very remarkable. So far, various PT-symmetric wave
systems have been investigated. Examples include the NLS equations
with linear and/or nonlinear PT-symmetric potentials (Musslimani
\emph{et al}. 2008, Wang \& Wang 2011, Lu \& Zhang 2011, Abdullaev
\emph{et al}. 2011, He \emph{et al}. 2012, Nixon \emph{et al}. 2012,
Yang 2012a, Zezyulin \& Konotop 2012a, Huang \emph{et al}. 2013),
vector NLS equations with PT-symmetric potentials (Kartashov 2013),
PT-symmetric couplers (Driben \& Malomed 2011, Alexeeva \emph{et
al}. 2012), the $\chi^{(2)}$ system with PT-symmetric potentials
(Moreira \emph{et al}. 2012), discrete NLS-PT lattices (Konotop
\emph{et al}. 2012, Kevrekidis \emph{et al}. 2013), and so on. In
these systems, continuous families of PT-symmetric solitary waves
(or solitons in short) have been reported. Other PT systems, such as
finite-dimensional PT systems, have also been explored (Li \&
Kevrekidis 2011, Zezyulin \& Konotop 2012b, Zezyulin \& Konotop
2013). In certain finite-dimensional PT systems (such as the
quadrimer model), it was found that continuous families of
PT-symmetric solutions could coexist with isolated solutions.
Experimentally, PT-symmetric potentials have been fabricated in
optical settings (Guo  \emph{et al}. 2009, R\"uter \emph{et al}.
2010).

Since solitons in PT systems could exist for continuous ranges of
propagation constants, bifurcations of such solitons become an
important issue. In conservative wave systems, various archetypical
bifurcations of solitons have been reported, including fold
bifurcations (also known as saddle-node or saddle-center
bifurcations), symmetry-breaking bifurcations (also known as
pitchfork bifurcations), and transcritical bifurcations, see Yang
(2012b) and the references therein. In PT systems, fold bifurcations
have been found (Yang 2012a, Zezyulin \& Konotop 2012a, Konotop
\emph{et al}. 2012, Kevrekidis \emph{et al}. 2013), but the other
types of bifurcations are still unknown.

Of all bifurcations, symmetry-breaking bifurcations are particularly
interesting, since such bifurcations create solitary waves that do
not obey the symmetry of the original system. One of the most
familiar symmetry-breaking bifurcations is in the NLS equation with
a real symmetric potential, where families of asymmetric solitons
bifurcate out from symmetric solitons at certain
propagation-constant values (Jackson \& Weinstein 2004, Kirr
\emph{et al}. 2008, Sacchetti 2009, Kirr \emph{et al}. 2011, Akylas
\emph{et al}. 2012, Pelinovsky \& Phan 2012, Yang 2013). These
asymmetric solitons are often more stable than their symmetric
counterparts. PT-symmetric systems have been shown to admit families
of PT-symmetric solitons. Then a natural question is, can symmetry
breaking occur for these PT-symmetric solitons? If so, continuous
families of non-PT-symmetric solitons would appear in a PT-symmetric
system.

Symmetry breaking is a dominant way for the creation of families of
asymmetric solitons, but it may not be the only way. Thus, a more
general question is, can PT-symmetric systems admit continuous
families of non-PT-symmetric solitons?

In this article, we investigate the existence of families of
non-PT-symmetric solitons in a familiar PT system --- the
one-dimensional NLS equation with a linear PT-symmetric potential.
This PT system governs paraxial nonlinear light propagation in a
medium with symmetric refractive index and anti-symmetric gain and
loss (Musslimani \emph{et al}. 2008, Yang 2010), as well as
Bose-Einstein condensates in a symmetric potential with balanced
gain and loss (Pitaevskii \& Stringari 2003). For this PT system, we
show that continuous families of asymmetric solitons in a real
symmetric potential are destroyed when this real potential is
perturbed by weak PT-symmetric perturbations. We further show that
in a general one-dimensional PT-symmetric potential, symmetry
breaking of PT-symmetric solitons cannot occur. Based on these
findings, we conjecture that one-dimensonal PT-symmetric potentials
cannot support continuous families of non-PT-symmetric solitons. In
other words, continuous families of solitons in one-dimensional
PT-symmetric potentials must be PT-symmetric.

\section{Preliminaries}
Our study of solitary waves in PT-symmetric systems is based on the
following one-dimensional nonlinear Schr\"odinger (1D NLS) equation
with a linear PT-symmetric potential
\begin{equation}  \label{e:U}
iU_t+U_{xx}-V(x)\Psi+\sigma |\Psi|^2\Psi=0,
\end{equation}
where $V(x)$ is a complex-valued (non-real) PT-symmetric potential
\[ V^*(x)=V(-x),   \label{e:PT} \]
with the asterisk representing complex conjugation, and $\sigma=\pm
1$ is the sign of nonlinearity ($\sigma=1$ for self-focusing and
$\sigma=-1$ for self-defocusing). Here, the PT-symmetry (\ref{e:PT})
means that the real part of the potential $V(x)$ is symmetric in
$x$, and the imaginary part of $V(x)$ is antisymmetric in $x$. The
equation (\ref{e:U}) is the appropriate mathematical model for
paraxial light transmission in PT-symmetric media (where the
refractive index is symmetric and gain-loss profile anti-symmetric).
It also governs the dynamics of Bose-Einstein condensates in a
symmetric potential with spatially-balanced gain and loss (in this
community, Eq. (\ref{e:U}) is called the Gross-Pitaevskii equation).
In the model (\ref{e:U}), the nonlinearity is only cubic. But
extension of our analysis to an arbitrary form of nonlinearity is
straightforward without much more effort (see Yang 2012b).

Solitary waves in Eq. \eqref{e:U} are sought of the form
\begin{equation}  \label{e:Usoliton}
U(x,t)=e^{i\mu t}u(x),
\end{equation}
where $u(x)$ is a complex-valued localized function which satisfies
the equation
\begin{equation}  \label{e:u}
u_{xx}-\mu u-V(x)u+\sigma |u|^2u=0,
\end{equation}
and $\mu$ is a real-valued propagation constant. Even though Eq.
\eqref{e:U} is dissipative due to the complex potential $V(x)$, a
remarkable phenomenon is that it can support continuous families of
solitary waves (\ref{e:Usoliton}), parameterized by the propagation
constant $\mu$ --- just like in real potentials (Wang \& Wang 2011,
Lu \& Zhang 2011, He \emph{et al}. 2012, Nixon \emph{et al}. 2012,
Yang 2012a, Zezyulin \& Konotop 2012a, Huang \emph{et al}. 2013).
Then under certain conditions, these solitary waves may undergo
bifurcations at special values of $\mu$.

If the potential $V(x)$ were strictly real, then the PT symmetry
condition (\ref{e:PT}) would become $V(-x)=V(x)$, i.e., this
potential would be symmetric. It is well known that in real
symmetric potentials, symmetry breaking of solitary waves often
occurs. Specifically, in addition to continuous families of
symmetric and anti-symmetric solitons, continuous families of
asymmetric solitons can also bifurcate out from those symmetric and
anti-symmetric soliton branches. This symmetry breaking is most
familiar in double-well potentials (Jackson \& Weinstein 2004,
Sacchetti 2009), but it can occur in other symmetric potentials
(such as periodic potentials) as well (Kirr \emph{et al.} 2008,
Akylas \emph{et al.} 2012). Due to this symmetry breaking,
continuous families of asymmetric solitons appear in a real
symmetric potential.

When the potential $V(x)$ is complex but PT-symmetric, there exist
continuous families of solitary waves $u(x; \mu)$ with the same
PT-symmetry
\[ u^*(x; \mu)=u(-x; \mu), \]
see Wang \& Wang (2011), Lu \& Zhang (2011), Nixon \emph{et al}.
(2012), Zezyulin \& Konotop (2012a). Then, the question is, can
PT-symmetry breaking occur for these PT-symmetric solitons? More
generally, can continuous families of non-PT-symmetric solitons
exist in a one-dimensional PT-symmetric potential? These are the
questions we will address in this article.

\textbf{Remark 1} \hspace{0.02cm} It is noted that Eq. (\ref{e:u})
is phase-invariant. That is, if $u(x)$ is a solitary wave, then so
is $u(x)e^{i\alpha}$, where $\alpha$ is any real constant. For a
solitary wave $u(x)$, if there exists a real constant $\alpha$ so
that $u(x)e^{i\alpha}$ is PT-symmetric, then we say $u(x)$ is
reducible to PT-symmetric. For instance, a complex solitary wave
$u(x)$ with anti-PT-symmetry $u^*(x)=-u(-x)$, i.e., with an
anti-symmetric real part and symmetric imaginary part, is reducible
to PT-symmetric by multiplying it by $i$. In general, a simple way
to determine whether a complex-valued solitary wave $u(x)$ is
reducible to PT-symmetric or not is to examine the function
$u(x)e^{-i\theta}$, where $\theta$ is the phase of $u(0)$. If $u(x)
e^{-i\theta}$ is PT-symmetric, then $u(x)$ is reducible to
PT-symmetric; and vise versa. Graphically, a simple way to decide
whether a solitary wave $u(x)$ is reducible to PT-symmetric is to
plot the amplitude $|u(x)|$ of the function. If this amplitude is
not symmetric in $x$, then $u(x)$ is not reducible to PT-symmetric.
In this article, when we say non-PT-symmetric solitary waves or
solitons, we mean solitary waves that are not reducible to
PT-symmetric.

\section{Disappearance of families of asymmetric solitons under weak PT-potential
perturbations}

To explore the existence of continuous families of non-PT-symmetric
solitons in Eq. (\ref{e:U}) with a PT-symmetric potential, we first
investigate what happens to families of asymmetric solitons of a
real symmetric potential when this real potential is weakly
perturbed by an imaginary anti-symmetric term (which makes the
perturbed potential non-real but PT-symmetric). Can these families
of asymmetric solitons survive and turn into families of
non-PT-symmetric solitons in the resulting PT potential? The answer
is negative. We will show that these families of asymmetric solitons
of real potentials disappear under weak PT-potential perturbations.

When a real symmetric potential is perturbed by an imaginary
anti-symmetric term, the model equation (\ref{e:u}) can be written
as
\begin{equation}  \label{e:up}
u_{xx}-\mu u-V_r(x)u+\sigma |u|^2u=i\epsilon W(x)u,
\end{equation}
where $V_r(x)$ is a real symmetric potential, $W(x)$ is a real
anti-symmetric function,
\[ V_r(x)=V_r(-x), \quad W(x)=-W(-x), \]
and $\epsilon$ is a small real parameter. Note that the combined
potential in Eq. (\ref{e:up}), $V=V_r+i\epsilon W$, is complex and
PT-symmetric.

Suppose we seek a continuous family of solitons in the perturbed
potential (with $0<\epsilon\ll 1$), near a continuous family of
asymmetric solitons in the unperturbed real potential (with
$\epsilon=0$). Since the soliton family in the perturbed potential
should exist for a continuous range of $\mu$ values, each soliton in
that family with a fixed $\mu$ value should converge to the
asymmetric soliton of $\epsilon=0$ with the same $\mu$ value when
$\epsilon\to 0$. Because of this, in our perturbation expansion we
can fix $\mu$ and expand the soliton at this $\mu$ value into a
perturbation series of $\epsilon$,
\[  \label{e:uexpand2}
u(x; \epsilon)=u_r(x)+\epsilon u_1(x)+ \epsilon^2 u_2(x)+\dots,
\]
where $u_r(x)$ is an asymmetric (i.e. neither symmetric nor
anti-symmetric) real soliton in the real potential $V_r$, i.e.,
\[  \label{e:u02}
\frac{d^2u_r}{dx^2}-\mu u_r-V_r(x)u_r+\sigma u_r^3=0,
\]
and
\[ \label{e:ura}
u_r(-x)\ne \pm u_r(x).
\]

We will show below that, in order for this perturbation series to be
constructed, an infinite number of non-trivial conditions would have
to be satisfied, which is impossible in practice. This conclusion
will also be corroborated by several specific examples.

We start by substituting the expansion (\ref{e:uexpand2}) into Eq.
(\ref{e:up}). The $O(1)$ equation is satisfied automatically due to
Eq. (\ref{e:u02}). At $O(\epsilon)$, the equation for $u_1$ is
\[  \label{e:u1b}
L_r \left(\begin{array}{c} u_1 \\ u_1^* \end{array}\right) =
\left(\begin{array}{c} iWu_r \\ -iWu_r  \end{array}\right),
\]
where
\[ \label{e:Lbdef}
L_r=\left[\begin{array}{cc} \frac{d^2}{dx^2} -V_r(x) -\mu +2\sigma u_r^2  & \sigma u_r^2 \\
\sigma u_r^{2} & \frac{d^2}{dx^2} -V_r(x) -\mu +2\sigma u_r^2 \end{array}\right]
\]
is a real and self-adjoint operator. The kernel of $L_r$ contains an
eigenfunction $[u_r, -u_r]^T$, where the superscript `\emph{T}'
represents the transpose of a vector, due to Eq. (\ref{e:u02}).
Thus,
\[
L_r  \left(\begin{array}{c} u_r \\ -u_r \end{array}\right) =0.
\]
Let us assume that the kernel of $L_r$ does not contain any
additional eigenfunctions, which is true for generic values of
$\mu$. Then the solvability condition for Eq. (\ref{e:u1b}) is that
its right hand side be orthogonal to $[u_r, -u_r]^T$, which reduces
to
\[  \label{e:cond2b}
Q_1(\mu)\equiv \langle u_r, Wu_r \rangle =0,
\]
where the inner product is defined as
\begin{equation} \label{def:inner_prod}
\langle f, g\rangle  = \int_{-\infty}^\infty f^{*T}(x) \hspace{0.05cm}
g(x) \hspace{0.07cm} d x.
\end{equation}
Since $W$ is anti-symmetric and $u_r$ asymmetric [see
(\ref{e:ura})], Eq. (\ref{e:cond2b}) is then a non-trivial condition
which must be satisfied in order for the asymmetric soliton $u_r(x)$
to persist under weak PT-potential perturbations.

It turns out that Eq. (\ref{e:cond2b}) is only the first of
infinitely many conditions which must be satisfied in order for the
perturbation-series solution (\ref{e:uexpand2}) to be constructed.
Indeed, at each higher odd order, a new condition would appear. For
instance, if condition (\ref{e:cond2b}) is met, then the $u_1$
equation (\ref{e:u1b}) can be solved. This solution can be written
as
\[
u_1=i\hat{u}_1+id_1u_r,
\]
where $\hat{u}_1$ is a real and localized function solving the
equation
\[  \label{e:u1bhat}
L_r \left(\begin{array}{c} \hat{u}_1 \\ -\hat{u}_1 \end{array}\right) =
\left(\begin{array}{c} Wu_r \\ -Wu_r  \end{array}\right),
\]
and $d_1$ is a real constant. This $id_1u_r$ term in $u_1$, when
combined with the first term $u_r$ in the expansion
(\ref{e:uexpand2}), only amounts to a constant phase shift to the
solution $u(x; \epsilon)$, thus it can be set to be zero without any
loss of generality (see Remark 1). Thus,
\[
u_1=i\hat{u}_1.
\]

At $O(\epsilon^2)$, the $u_2$ equation is
\[
L_r \left(\begin{array}{c} u_2 \\ u_2^* \end{array}\right) =
\left(\begin{array}{c} h_2 \\ h_2  \end{array}\right),
\]
where
\[
h_2=-\sigma u_r\hat{u}_1^2-W\hat{u}_1
\]
is a real function. The solvability condition for this equation is
satisfied automatically, thus $u_2$ has a solution
\[
u_2=\hat{u}_2,
\]
where $\hat{u}_2$ is a real and localized function. As before, we
have excluded the homogeneous term (proportional to $iu_r$) in the
$u_2$ solution without loss of generality.

Now we proceed to $O(\epsilon^3)$, where the $u_3$ equation is
\[
L_r \left(\begin{array}{c} u_3 \\ u_3^* \end{array}\right) =
\left(\begin{array}{c} ih_3 \\ -ih_3  \end{array}\right),
\]
and
\[
h_3=-\sigma(\hat{u}_1^3+2u_r\hat{u}_1\hat{u}_2)+W\hat{u}_2
\]
is a real function. The solvability condition of this $u_3$ equation
is that
\[  \label{e:cond3b}
Q_2(\mu) \equiv \langle u_r, h_3 \rangle =0.
\]
Since $u_r, \hat{u}_1$ and $\hat{u}_2$ are all asymmetric functions,
so is $h_3$. Then, Eq. (\ref{e:cond3b}) is the second non-trivial
condition which has to be met. Following similar calculations to
higher orders, infinitely more conditions will appear.

The fundamental reason for this infinite number of conditions for
the perturbation series solution (\ref{e:uexpand2}) is that, due to
phase invariance of the solitons in Remark 1, each $u_n$ solution
does not contain any non-reducible free constants. But for each odd
$n$, the $u_n$ equation is of the form $L_r[u_n, u_n^*]^T=[ih_n,
-ih_n]^T$, where $h_n$ is a certain real function. In order for this
$u_n$ equation to be solvable, the solvability condition $\langle
u_r, h_n\rangle=0$ must be satisfied. All these solvability
conditions then constitute an infinite number of conditions for the
perturbation series solution (\ref{e:uexpand2}).

In one spatial dimension, neither the perturbed PT-symmetric
potential nor the underlying asymmetric solitons possesses
additional spatial symmetries. Because of that, each of these
infinitely many conditions is non-trivial and is generally not
satisfied for generic values of $\mu$. The requirement of them all
satisfied simultaneously is practically impossible. This means that
continuous families of asymmetric solitons in the real potential
would disappear under weak PT-potential perturbations.

Now we use three specific examples to corroborate the above
statement.

\textbf{Example 1.} \ In this first example, we take $V_r$ to be a
symmetric double-well potential
\[  \label{e:realpot}
V_r(x)=-3\left[\mbox{sech}^2(x+1.5)+\mbox{sech}^2(x-1.5)\right],
\]
and $W$ to be an anti-symmetric function
\[  \label{e:Wx}
W(x)=\mbox{sech}^2(x+1.5)-\mbox{sech}^2(x-1.5).
\]
Both functions are displayed in Fig. 1(a). In addition, we take
$\sigma=1$, i.e., self-focusing nonlinearity.

In this double-well potential $V_r$, a branch of real symmetric
solitons exist, whose power curve is shown in Fig. 1(b) (solid blue
line). In addition, symmetry breaking occurs at $\mu_0\approx
2.1153$, where a branch of asymmetric solitons appear for
$\mu>\mu_0$. An example of such asymmetric solitons (with $\mu=2.3$)
is illustrated in Fig. 1(c). For these asymmetric solitons, we have
numerically calculated the function $Q_1(\mu)$ as defined in Eq.
(\ref{e:cond2b}), and this function is plotted in Fig. 1(d). We can
see that this function is non-zero for all $\mu>\mu_0$, thus the
first condition (\ref{e:cond2b}) is never satisfied, let alone all
the other conditions such as (\ref{e:cond3b}). Thus we conclude that
in this example, the continuous family of asymmetric solitons in the
real symmetric potential (\ref{e:realpot}) are destroyed under weak
PT-potential perturbations (\ref{e:Wx}).

\begin{figure}[tb!]
\centerline{
\includegraphics[width=0.8\textwidth]{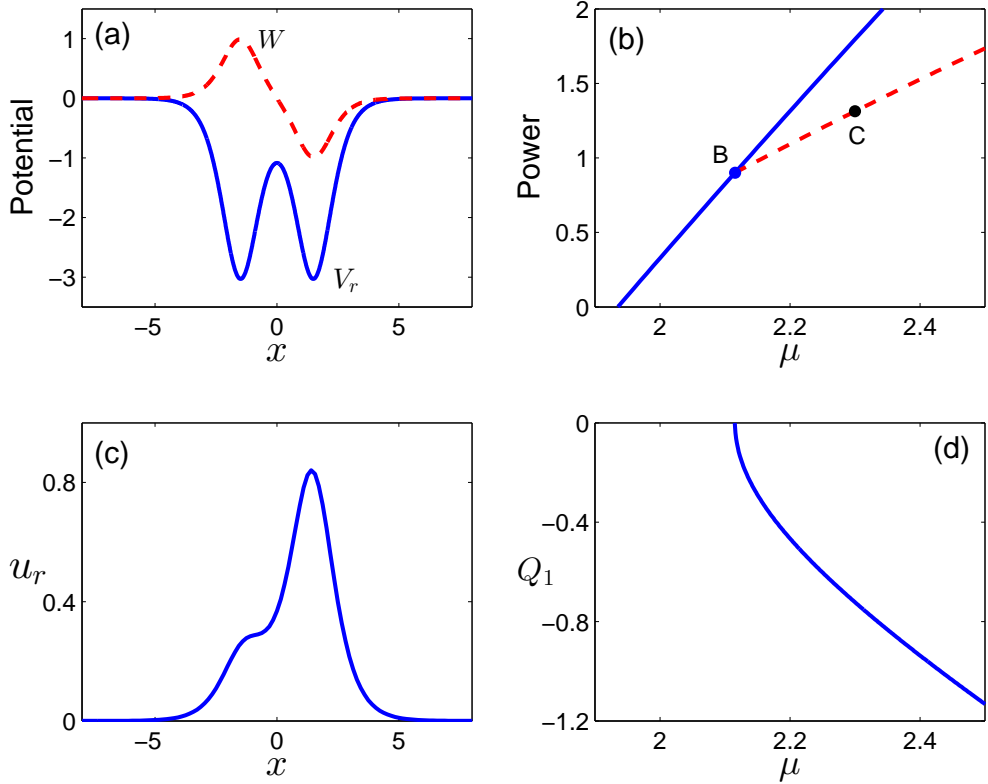}
}
\smallskip
\caption{(Color online) Disappearance of the continuous family of asymmetric solitons under PT-potential perturbations in Example 1.
(a) Real potential (\ref{e:realpot}) and PT perturbation (\ref{e:Wx});
(b) power curves of symmetric (blue solid) and asymmetric (red
dashed) solitons; (c) asymmetric soliton $u_r(x)$ at point `C' of panel (b), where $\mu=2.3$;
(d) function $Q_1(\mu)$ in condition (\ref{e:cond2b}). This function being non-zero indicates that the continuous family of asymmetric solitons
(red dashed line in panel (b)) are destroyed under PT-potential perturbations. } \label{fig1}
\end{figure}

\textbf{Example 2.} \ In the second example, we keep the real
potential $V_r$ of Example 1, but choose a different anti-symmetric
function for $W$ as
\[ \label{e:Wx2}
W(x)=\mbox{sech}^2(x+1.5)\: \mbox{tanh}(x+1.5)+\mbox{sech}^2(x-1.5)\: \mbox{tanh}(x-1.5).
\]
This new function $W$ is displayed in Fig. 2(a). The significance of
this new $W$ function is that it is proportional to $V'_r(x)$. In
this case, multiplying Eq. (\ref{e:u02}) by $u_r'(x)$ and
integrating from $-\infty$ to $+\infty$, we find that
\[
\langle u_r, V'_ru_r \rangle=0.  \nonumber
\]
Since $W\propto V'_r$, $Q_1(\mu)$ is then always zero, thus the
first condition (\ref{e:cond2b}) is satisfied automatically for all
$\mu>\mu_0$. However, for this $W$ function, the second condition
(\ref{e:cond3b}) is never satisfied. Indeed, we have numerically
computed the function $Q_2$ in this condition and plotted it in Fig.
2(b); one can see that it is never zero for $\mu>\mu_0$. Since this
second condition is not met, this family of asymmetric solitons
cannot persist and have to disappear under PT-potential
perturbations (\ref{e:Wx2}) as well.

\begin{figure}[tb!]
\centerline{
\includegraphics[width=0.8\textwidth]{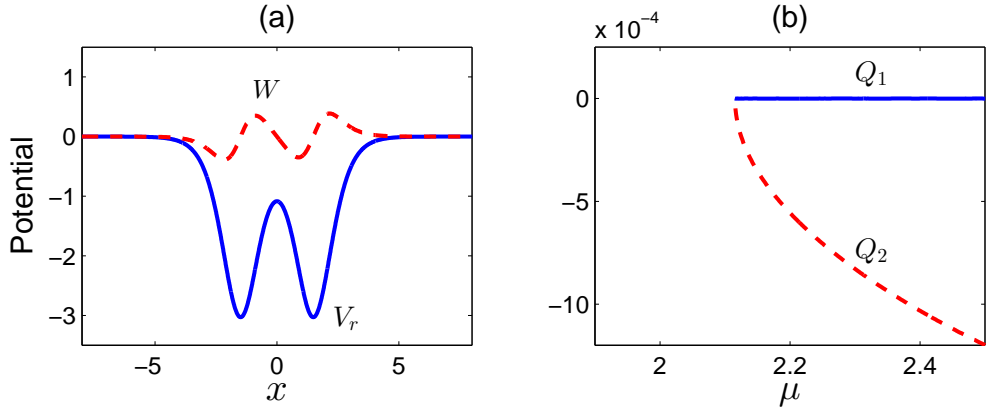}
}
\smallskip
\caption{(Color online) Disappearance of the continuous family of asymmetric solitons under PT-potential perturbations in Example 2.
(a) Real potential (\ref{e:realpot}) and PT perturbation (\ref{e:Wx2});
(b) functions $Q_1(\mu)$ and $Q_2(\mu)$ in conditions (\ref{e:cond2b}) and (\ref{e:cond3b}).
Here $Q_2\ne 0$, thus the continuous family of asymmetric solitons are destroyed under
PT-potential perturbations. } \label{fig2}
\end{figure}

\textbf{Example 3.} \ In the third example, we keep the real
potential $V_r$ of Examples 1 and 2, but choose yet another
anti-symmetric function for $W$ as
\[ \label{e:Wx3}
W(x)=\mbox{sech}^2(x+1.5)-\mbox{sech}^2(x-1.5) - 1.15 \left[\mbox{sech}^2(x+2)-\mbox{sech}^2(x-2)\right].
\]
This $W$ function is plotted in Fig. 3(a). For this choice of $W$,
the function $Q_1(\mu)$ in condition (\ref{e:cond2b}) is displayed
in Fig. 3(b). We see that this $Q_1$ is zero only at a special $\mu$
value of $\mu_c \approx  2.5343$, which is marked as a red dot in
Fig. 3(b). Because of that, under this $W(x)$ perturbation, the
continuous family of asymmetric solitons (with $\mu\ne \mu_c$) in
the real potential (\ref{e:realpot}) are all destroyed.

\begin{figure}[tb!]
\centerline{
\includegraphics[width=0.8\textwidth]{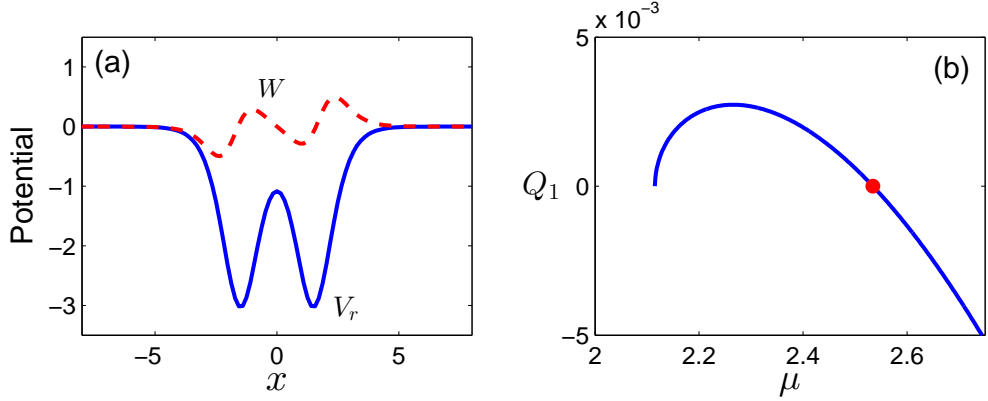}
}
\smallskip
\caption{(Color online) Disappearance of the continuous family of asymmetric solitons under PT-potential perturbations in Example 3.
(a) Real potential (\ref{e:realpot}) and PT perturbation (\ref{e:Wx3});
(b) function $Q_1(\mu)$ in condition (\ref{e:cond2b}); this function is zero at the red-dot point.} \label{fig3}
\end{figure}

What about families of symmetric and anti-symmetric solitons of real
potentials under weak PT perturbations? In that case, repeating the
above perturbation calculations, we can easily show that all
conditions, such as (\ref{e:cond2b}) and (\ref{e:cond3b}), are
automatically satisfied due to symmetries of the involved functions.
As a consequence, perturbation series (\ref{e:uexpand2}) for
solitary waves $u(x; \epsilon)$ can be constructed to all orders. In
addition, the constructed solutions $u(x; \epsilon)$ are
PT-symmetric or reducible to PT-symmetric. This indicates that,
families of symmetric and anti-symmetric solitons of real
potentials, under PT perturbations, persist and turn into families
of PT-symmetric solitons.

\section{No symmetry breaking in PT-symmetric potentials}

In this section, we turn our attention to general PT-symmetric
potentials whose imaginary parts are not necessarily small. In such
a general PT potential, if Eq. (\ref{e:u}) admits a branch of
PT-symmetric solitons (which is often the case), we ask whether a
PT-symmetry-breaking bifurcation can occur, where new branches of
non-PT-symmetric solitons bifurcate out from this PT-symmetric
branch.

If the potential were real symmetric, symmetry breaking of solitary
waves would occur frequently (Jackson \& Weinstein 2004, Kirr
\emph{et al.} 2008, Sacchetti 2009, Akylas \emph{et al.} 2012).
However, when the potential becomes complex and PT-symmetric, we
will show that PT-symmetry breaking cannot occur in Eq. (\ref{e:u}).

Before the analysis, we first introduce some notations and make some
basic observations.

\subsection{Notations and simple observations}

The linearization operator of the solitary-wave equation (\ref{e:u})
plays an important role in the bifurcation analysis. Since the
solitary wave $u(x)$ is complex-valued due to the complex potential,
this linearization operator is vector rather than scalar and can be
written as
\[ \label{e:Ldef}
L=\left[\begin{array}{cc} \frac{d^2}{dx^2} -V(x) -\mu +2\sigma|u|^2  & \sigma u^2 \\
\sigma u^{*2} & \frac{d^2}{dx^2} -V^*(x) -\mu +2\sigma |u|^2 \end{array}\right].
\]
This operator is non-Hermitian. Under the standard inner product
(\ref{def:inner_prod}), the adjoint operator of $L$ is then
\[
L^A=\left[\begin{array}{cc} \frac{d^2}{dx^2} -V^*(x) -\mu +2\sigma |u|^2  & \sigma u^2 \\
\sigma u^{*2} & \frac{d^2}{dx^2} -V(x) -\mu +2\sigma |u|^2 \end{array}\right].
\]

The kernel of the linearization operator $L$ is clearly not empty.
Indeed, it is easy to see that
\[  \label{e:Lu}
L\left[\begin{array}{c} u \\ -u^* \end{array}\right]=0
\]
for all $\mu$ values in view of Eq. (\ref{e:u}), thus the dimension
of the kernel of $L$ is at least one.

For any eigenfunction $[f, g]^T$ in the kernel of $L$, it is easy to
see that $[g^*, f^*]^T$ is also in this kernel. By adding these two
eigenfunctions, we get an eigenfunction in the form of $[w, w^*]^T$,
or equivalently $[\hat{w}, -\hat{w}^*]^T$ if one sets $w=i\hat{w}$.
Eigenfunctions in these special forms will be chosen as the basis to
span the kernel of $L$. Similar statements go to the kernel of the
adjoint operator $L^A$ as well.

Our basic observation on solitary-wave bifurcations in Eq.
(\ref{e:u}) is that, if a bifurcation occurs at $\mu=\mu_0$, by
denoting the corresponding solitary wave and the linearization
operator as
\begin{equation}  \label{e:defu0}
u_0(x) \equiv u(x; \mu_0), \quad L_{0} \equiv  L|_{\mu=\mu_0,\ u=u_0},
\end{equation}
then dimension of the kernel of $L_{0}$ should be at least two,
i.e., dim[ker($L_0$)] $\ge 2$. This means that the kernel of $L_{0}$
should contain at least another localized eigenfunction in addition
to $[u_0, -u_0^*]^T$. Using the language of multiplicity of
eigenvalues, this means that zero should be a discrete eigenvalue of
$L_0$ with geometric multiplicity at least two. This is a necessary
(but not sufficient) condition for bifurcations.

The above necessary condition for bifurcations can be made even more
explicit. Since $L_0$ is a fourth-order ordinary differential
operator, and the Wronskian of fundamental solutions in its kernel
is a non-zero constant (see Remark 2 below), this kernel then cannot
contain more than two linearly independent localized eigenfunctions.
In other words, dim[ker($L_0$)] $\le 2$. Then, combined with the
above observation, we see that \emph{a necessary condition for
solitary-wave bifurcations in Eq. (\ref{e:u}) is that}
\[
\mbox{dim[ker($L_0$)]=2. }
\]

Due to this condition, if a bifurcation occurs at $\mu=\mu_0$, then
the kernel of $L_0$ would contain exactly one additional
eigenfunction, which can be denoted as $[\psi, \psi^*]^T$. Thus,
\[   \label{e:kernelL0}
L_0\left[\begin{array}{c} u_0 \\ -u_0^* \end{array}\right]=
L_0\left[\begin{array}{c} \psi \\ \psi^* \end{array}\right]=0.
\]

Since operator $L_0$ for PT potentials is non-self-adjoint, the
kernel of the adjoint operator $L_0^A$ will also play an important
role in the bifurcation analysis. When dim[ker($L_0$)]=2, we will
show in Remark 2 below that
\[
\mbox{dim[ker($L_0^A$)]=2 }
\]
as well. Thus the kernel of $L_0^A$ also contains exactly two
linearly independent localized eigenfunctions: one can be denoted as
$[\phi_1, -\phi_1^*]^T$, and the other denoted as $[\phi_2,
\phi_2^*]^T$, i.e.,
\[  \label{e:kernelL0A}
L_0^A\left[\begin{array}{c} \phi_1 \\ -\phi_1^* \end{array}\right]=
L_0^A\left[\begin{array}{c} \phi_2 \\ \phi_2^* \end{array}\right]=0.
\]

\textbf{Remark 2} \hspace{0.06cm} Here we show that if
dim[ker($L_0$)]=2, then dim[ker($L_0^A$)]=2. To show this, notice
that both ordinary differential operators $L_0$ and $L_0^A$ are
four-dimensional. Let us rewrite the operator equations $L_0Y=0$ and
$L_0^A Y^A=0$ as systems of four first-order equations, $Z_x=QZ$ and
$-Z^A_x=Q^\dagger Z^A$, where $Z=[Y_1, Y_{1x}, Y_2, Y_{2x}]^T$,
$Z^A=[Y_1^A, Y_{1x}^A, Y_2^A, Y_{2x}^A]^T$, and the superscript
`$\dagger$' represents Hermitian (i.e., transpose conjugation). Then
it is easy to see that $\mbox{tr}(Q)=0$, thus Wronskians of
fundamental matrices for these two first-order systems are both
non-zero constants. It is also known that if the fundamental matrix
of the system $Z_x=QZ$ is $Z=M$, then the fundamental matrix of the
adjoint system $-Z^A_x=Q^\dagger Z^A$ would be
$Z^A=(M^{-1})^\dagger$. Under the assumption of dim[ker($L_0$)]=2,
two columns of the fundamental matrix $M$ are localized functions.
Since the determinant of $M$ is a non-zero constant, using the
$M^{-1}$ formula in terms of cofactors, the other two columns in the
adjoint fundamental matrix $(M^{-1})^\dagger$ then are localized
functions. Hence, dimension of the kernel of $L_0^A$ is also two.

\vspace{0.2cm} \textbf{Remark 3} \hspace{0.06cm} If the solitary
wave $u_0(x)$ is PT-symmetric, then eigenfunctions in Eqs.
(\ref{e:kernelL0}) and (\ref{e:kernelL0A}) can always be chosen so
that they are either PT-symmetric or anti-PT-symmetric, i.e.,
\[ \label{e:kernelsym}
\psi^*(x)=\pm \psi(-x), \quad \phi_1^*(x)=\phi_1(-x), \quad  \phi_2^*(x)=\pm \phi_2(-x).
\]
To prove this, we notice that since $V(x)$ and $u_0(x)$ are both
PT-symmetric, by taking the complex conjugate of the second equation
in (\ref{e:kernelL0}) and switching $x$ to $-x$, then $L_0$ is
invariant, and $[\psi^*(-x), \psi(-x)]^T$ is also in the kernel of
$L_0$. Since the kernel of $L_0$ has dimension two, $[\psi^*(-x),
\psi(-x)]^T$ then should be a linear combination of $[u_0(x),
-u_0^*(x)]^T$ and $[\psi(x), \psi^*(x)]^T$, i.e.,
\[  \label{e:c1c2a}
\psi^*(-x)=c_1\psi(x)+c_2u_0(x),
\]
and
\[  \label{e:c1c2b}
\psi(-x)=c_1\psi^*(x)-c_2u_0^*(x),
\]
where $c_1$, $c_2$ are certain complex constants. Switching $x$ to
$-x$ in (\ref{e:c1c2b}) and using the PT symmetry of $u_0$, we get
\[  \label{e:c1c2c}
\psi(x)=c_1\psi^*(-x)-c_2u_0(x).
\]
Then adding (\ref{e:c1c2a}) and  (\ref{e:c1c2c}), we get
\[
(1-c_1) \left[ \psi(x)+\psi^*(-x)\right]=0.
\]
Thus, if $c_1\ne 1$, then $\psi^*(x)=-\psi(-x)$, i.e., $\psi(x)$ is
anti-PT-symmetric. If $c_1=1$, by taking the complex conjugate of
Eq. (\ref{e:c1c2b}) and then subtracting it from Eq.
(\ref{e:c1c2a}), we get $c_2^*=-c_2$, i.e., $c_2$ is purely
imaginary. Denoting $c_2=i\beta$, where $\beta$ is a real parameter,
Eq. (\ref{e:c1c2c}) can be rewritten as
\[  \label{e:psihat}
\hat{\psi}^*(x)=\hat{\psi}(-x),
\]
where $\hat{\psi}\equiv \psi+\frac{1}{2}i\beta u_0$. It is easy to
see that $[\hat{\psi}, \hat{\psi}^*]^T$ is a linear combination of
$[\psi(x), \psi^*(x)]^T$ and $[u_0(x), -u_0^*(x)]^T$, hence it is
also in the kernel of $L_0$. Then, instead of $[\psi(x),
\psi^*(x)]^T$, we can choose $[\hat{\psi}, \hat{\psi}^*]^T$  in the
eigenvalue equation (\ref{e:kernelL0}); and now $\hat{\psi}$ is
PT-symmetric in view of Eq. (\ref{e:psihat}). Symmetries
(\ref{e:kernelsym}) for kernels of $L_0^A$ in (\ref{e:kernelL0A})
can be proved in a similar way.

\subsection{Nonexistence of PT-symmetry breaking}

The observations and remarks in the previous subsection apply to all
bifurcations in the PT system (\ref{e:u}). Now we focus on the
particular type of bifurcation: symmetry-breaking bifurcation.

Suppose $u_s(x; \mu)$ is a base branch of PT-symmetric solitons. If
a symmetry-breaking bifurcation occurs at $\mu=\mu_0$ of this base
branch, with $u_0(x) \equiv u_s(x; \mu_0)$, then eigenfunctions
(\ref{e:kernelL0}) and (\ref{e:kernelL0A}) in the kernels of $L_0$
and $L_0^A$ should have the following symmetries
\[  \label{e:sym1}
u_0^*(x)=u_0(-x), \quad \psi^*(x)=-\psi(-x),
\]
\[ \label{e:sym2}
\phi_1^*(x)=\phi_1(-x), \quad
\phi_2^*(x)=-\phi_2(-x),
\]
i.e., $u_0$, $\phi_1$ are PT-symmetric, and $\psi$, $\phi_2$
anti-PT-symmetric (see Remark 3). In addition, since the two
functions in the kernel of $L_0$ should be linearly independent,
$\psi\ne iu_0$. For a similar reason, $\phi_2\ne i\phi_1$.

Below we will show that, in a general PT-symmetric potential, the
kernel of $L_0$ generically cannot contain the second eigenfunction
$[\psi, \psi^*]^T$ with anti-PT-symmetry (\ref{e:sym1}). Thus the
necessary condition for symmetry breaking is not met. We will also
show that even if such a second eigenfunction $[\psi, \psi^*]^T$
appears in the kernel of $L_0$, symmetry breaking still cannot
occur.

First, we show that in a general PT-symmetric potential, the kernel
of $L_0$ generically cannot contain the second eigenfunction $[\psi,
\psi^*]^T$ with anti-PT-symmetry (\ref{e:sym1}). Using the language
of multiplicity of eigenvalues, we will show that when the zero
eigenvalue of $L_0$ has algebraic multiplicity higher than one, its
geometric multiplicity generically cannot be higher than one with an
anti-PT-symmetric second eigenfunction $[\psi, \psi^*]^T$.

Suppose when $\mu=\mu_0$, the zero eigenvalue of $L_0$ has algebraic
multiplicity higher than one and geometric multiplicity two, and the
second eigenfunction $[\psi, \psi^*]^T$ of this zero eigenvalue is
anti-PT-symmetric. When $\mu\ne \mu_0$, the zero eigenvalue of this
$[\psi, \psi^*]^T$ eigenmode would move out of the origin. Let us
calculate this eigenvalue of $L$ for $|\mu-\mu_0|\ll 1$ by
perturbation methods. The eigenvalue equation is
\[ \label{e:Lw}
L\left[\begin{array}{c} w \\ w^* \end{array}\right]=\lambda \left[\begin{array}{c} w \\ w^* \end{array}\right].
\]
When $|\mu-\mu_0|\ll 1$, we can expand the eigenvalue $\lambda$ and
the eigenfunction $[w, w^*]^T$ into a perturbation series,
\[
\lambda=\lambda_1 (\mu-\mu_0) + \lambda_2 (\mu-\mu_0)^2  +\dots,
\]
\[ \label{e:wexpand}
\left[\begin{array}{c} w \\ w^* \end{array}\right]= \left[\begin{array}{c} \psi \\ \psi^* \end{array}\right] +
(\mu-\mu_0)\left[\begin{array}{c} w_1 \\ w_1^* \end{array}\right] + (\mu-\mu_0)^2 \left[\begin{array}{c} w_2 \\ w_2^* \end{array}\right] +\dots.
\]
Similarly, we also expand the operator $L$ into a perturbation
series,
\[
L=L_0+(\mu-\mu_0)L_1+(\mu-\mu_0)L_2+\dots.
\]
When this eigenmode $[w, w^*]^T$ moves out of the origin, using
similar arguments as in Remark 3, we can show that $w$ can be made
PT-symmetric or anti-PT-symmetric. Since $w\to \psi$ as $\mu\to
\mu_0$ and $\psi$ is anti-PT-symmetric, $w$ then should be
anti-PT-symmetric. As a consequence, the other functions $w_1, w_2,
\dots$ in the $w$ expansion are also anti-PT-symmetric.

Substituting the above expansions into Eq. (\ref{e:Lw}), the $O(1)$
equation is satisfied automatically since $[\psi, \psi^*]^T$ is in
the kernel of $L_0$ by assumption. At $O(\mu-\mu_0)$, we get
\[  \label{e:w1}
L_0 \left[\begin{array}{c} w_1 \\ w_1^* \end{array}\right] = \lambda_1 \left[\begin{array}{c} \psi \\ \psi^* \end{array}\right]-
\left[\begin{array}{c} g_1 \\ g_1^* \end{array}\right],
\]
where
\[
\left[\begin{array}{c} g_1 \\ g_1^* \end{array}\right] \equiv
L_1 \left[\begin{array}{c} \psi \\ \psi^* \end{array}\right].
\]
The function $g_1$ is anti-PT-symmetric in view that $L_1$ is
PT-symmetric and $\psi$ anti-PT-symmetric.

Since the kernel of $L_0$ has dimension two under the current
assumption, the kernel of $L_0^A$ has dimension two as well (see
Remark 2), and the two linearly independent eigenfunctions in the
kernel of $L_0^A$ are denoted in Eq. (\ref{e:kernelL0A}) with
symmetries (\ref{e:sym2}). Then in order for the $w_1$ equation
(\ref{e:w1}) to be solvable, the solvability condition is that the
right side of (\ref{e:w1}) be orthogonal to the two eigenfunctions
in the kernel of $L_0^A$, i.e.,
\[ \label{e:solva1}
\left\langle \left[\begin{array}{c} \phi_1 \\ -\phi_1^* \end{array}\right], \hspace{0.3cm}
\lambda_1 \left[\begin{array}{c} \psi \\ \psi^* \end{array}\right]-
\left[\begin{array}{c} g_1 \\ g_1^* \end{array}\right] \right\rangle=0,
\]
\[ \label{e:solva2}
\left\langle \left[\begin{array}{c} \phi_2 \\ \phi_2^* \end{array}\right], \hspace{0.3cm}
\lambda_1 \left[\begin{array}{c} \psi \\ \psi^* \end{array}\right]-
\left[\begin{array}{c} g_1 \\ g_1^* \end{array}\right] \right\rangle=0.
\]
These two conditions give two different expressions for the same
eigenvalue coefficient $\lambda_1$. In order for these two formulae
to be consistent, the following compatibility condition must be
satisfied,
\[  \label{e:cond1}
\frac{\mbox{Im}\langle \phi_1, g_1\rangle}{\mbox{Im}\langle \phi_1, \psi\rangle}=
\frac{\mbox{Re}\langle \phi_2, g_1\rangle}{\mbox{Re}\langle \phi_2, \psi\rangle}.
\]
Here `Re' and `Im' represent the real and imaginary parts of a
complex number. Recalling the PT symmetry of $\phi_1$ and
anti-PT-symmetries of $\phi_2, \psi$ and $g_1$, $\langle \phi_1,
g_1\rangle$ and $\langle \phi_1, \psi\rangle$ are purely imaginary,
and $\langle \phi_2, g_1\rangle$, $\langle \phi_2, \psi\rangle$ are
strictly real. In addition, recalling that $\phi_2 \ne i\phi_1$, Eq.
(\ref{e:cond1}) then is a non-trivial compatibility condition for
the existence of a second eigenfunction $[\psi, \psi^*]^T$ in the
kernel of $L_0$. Since this compatibility condition is not satisfied
generically, the necessary condition for symmetry breaking is then
not met.

Next, we show that even if such a second eigenfunction $[\psi,
\psi^*]^T$ appears in the kernel of $L_0$, symmetry breaking still
cannot occur, because such a bifurcation further requires an
infinite number of additional non-trivial conditions to be satisfied
simultaneously, which is impossible in practice.

Suppose at a propagation constant $\mu=\mu_0$, the kernels of $L_0$
and $L_0^A$ have dimension two, and their eigenfunctions are given
in Eqs. (\ref{e:kernelL0}) and (\ref{e:kernelL0A}) with symmetries
(\ref{e:sym1}) and (\ref{e:sym2}). If a symmetry-breaking
bifurcation occurs at this point, then two new branches of
non-PT-symmetric solitons would bifurcate out from $u_0(x)$ on only
one side of $\mu=\mu_0$. Let us seek such non-PT-symmetric solitons
near $\mu=\mu_0$ by perturbation methods.

Suppose these new solitons bifurcate to the right side of $\mu_0$,
then their perturbation series can be written as
\[  \label{e:uaseries}
u_a(x; \mu)=\sum_{k=0}^\infty (\mu-\mu_0)^{k/2} u_k(x).
\]
Substituting this perturbation series into Eq. (\ref{e:u}), the
$O(1)$ equation is satisfied automatically since $u_0$ is a solitary
wave at $\mu=\mu_0$. At $O[(\mu-\mu_0)^{1/2}]$, the equation for
$u_1$ is
\[
L_0 \left(\begin{array}{c} u_1 \\ u_1^* \end{array}\right) = 0.
\]
In view of the kernel structure of $L_0$ in Eq. (\ref{e:kernelL0}),
we see that
\[  \label{e:u1u1star}
\left(\begin{array}{c} u_1 \\ u_1^* \end{array}\right) =c_1
\left(\begin{array}{c} \psi \\ \psi^* \end{array}\right)+d_1
\left(\begin{array}{c} u_0 \\ -u_0^* \end{array}\right),
\]
where $c_1$, $d_1$ are constants. In order for the resulting $u_1^*$
formula to be complex conjugate of the $u_1$ formula, $c_1$ must be
strictly real, and $d_1$ purely imaginary. Then the $d_1u_0$ term in
$u_1$, when combined with the leading-order term $u_0$ in the
expansion (\ref{e:uaseries}), only amounts to a phase shift to
$u_a(x; \mu)$, which is insignificant in view of Remark 1. Thus we
can set $d_1=0$ without loss of generality. Then the $u_1$ solution
becomes
\[  \label{e:u1sol}
u_1=c_1 \psi,
\]
where $c_1$ is a real constant.

For symmetry-breaking bifurcation to occur, $c_1$ should be
non-zero. In this case, the first-two-term solution of
(\ref{e:uaseries}),
$$u_0+(\mu-\mu_0)^{1/2} c_1\psi,$$
is not PT-symmetric, nor is it reducible to PT-symmetric, because
$u_0$ is PT-symmetric but $\psi$ is anti-PT-symmetric and $\psi \ne
iu_0$. This non-PT-symmetry will not be affected by higher-order
terms of (\ref{e:uaseries}), thus the resulting solution $u_a(x;
\mu)$ in (\ref{e:uaseries}) would be non-PT-symmetric.

At $O(\mu-\mu_0)$, the equation for $u_2$ is
\[  \label{e:u2}
L_0  \left(\begin{array}{c} u_2 \\ u_2^* \end{array}\right) =
\left(\begin{array}{c} g_2 \\ g_2^* \end{array}\right),
\]
where
\[ \label{e:g2}
g_2=u_0-\sigma c_1^2 (2u_0|\psi|^2+u_0^*\psi^2).
\]
Here the $u_1$ solution (\ref{e:u1sol}) has been utilized. Since
$u_0$ is PT-symmetric and $\psi$ anti-PT-symmetric, $g_2$ is
PT-symmetric.

The solvability conditions of Eq. (\ref{e:u2}) are that its right
hand side be orthogonal to the kernels of $L_0^A$ in Eq.
(\ref{e:kernelL0A}), i.e.,
\[  \label{e:sol1}
\mbox{Im} \langle \phi_1, g_2 \rangle = \mbox{Re} \langle \phi_2, g_2 \rangle =0.
\]
Recalling the symmetries of $\phi_1$ and $\phi_2$ in (\ref{e:sym2})
as well as the PT-symmetry of $g_2$, we see that $\langle \phi_1,
g_2 \rangle$ is strictly real, and $\langle \phi_2, g_2 \rangle$ is
purely imaginary, thus both solvability conditions in Eq.
(\ref{e:sol1}) are automatically satisfied. As a result, a localized
particular solution $\hat{u}_2$ can be found. This particular
solution can be split into two parts, corresponding to the two terms
of $g_2$ in (\ref{e:g2}):
\[
\hat{u}_2=\hat{u}_{21}+c_1^2 \hat{u}_{22}.
\]
Here, $\hat{u}_{21}$ solves
\[
L_0 \left(\begin{array}{c} \hat{u}_{21} \\ \hat{u}_{21}^* \end{array}\right) =\left(\begin{array}{c} u_0 \\ u_0^* \end{array}\right),
\]
and $\hat{u}_{22}$ solves
\[
L_0 \left(\begin{array}{c} \hat{u}_{22} \\ \hat{u}_{22}^* \end{array}\right) =\left(\begin{array}{c} -\sigma (2u_0|\psi|^2+u_0^*\psi^2) \\  -\sigma (2u_0|\psi|^2+u_0^*\psi^2)^*
\end{array}\right).
\]
Since both terms of $g_2$ are PT-symmetric, $\hat{u}_{21}$ and
$\hat{u}_{22}$ can be made PT-symmetric as well. The general
solution of $u_2$ is then this particular solution plus the
homogeneous solutions. Similar to the $u_1$ solution case, we can
exclude the homogeneous $u_0$ term and set
\[  \label{e:u2sol}
u_2=\hat{u}_{21}+c_1^2 \hat{u}_{22}+c_2\psi
\]
without loss of generality. Here $c_2$ is another real constant to
be determined.

The calculations so far have been benign. However, from the next
order, we will start to get an infinite number of additional
conditions which have to be satisfied in order for the perturbation
series (\ref{e:uaseries}) to be constructed. Let us begin with the
$u_3$ equation, which is
\[  \label{e:u3}
L_0\left(\begin{array}{c} u_3 \\ u_3^* \end{array}\right) =\left(\begin{array}{c} g_3 \\ g_3^* \end{array}\right),
\]
where
\[
g_3=u_1-\sigma\left(2u_0^*u_1u_2+2u_0u_1^*u_2+2u_0u_1u_2^*+|u_1|^2u_1\right).
\]
Substituting the $u_1$ and $u_2$ solutions (\ref{e:u1sol}) and
(\ref{e:u2sol}) into $g_3$, we get
\[  \label{e:g3}
g_3=c_1\left(g_{31}+c_1^2g_{32}+c_2g_{33}\right),
\]
where
\[
g_{31}=\psi-2\sigma \left(u_0^*\psi\hat{u}_{21}+u_0\psi^*\hat{u}_{21}+u_0\psi\hat{u}_{21}^*\right),   \nonumber
\]
\[
g_{32}=-2\sigma \left(u_0^*\psi\hat{u}_{22}+u_0\psi^*\hat{u}_{22}+u_0\psi\hat{u}_{22}^*\right)-\sigma |\psi|^2\psi,  \nonumber
\]
and
\[
g_{33}=-2\sigma (2u_0|\psi|^2+u_0^*\psi^2).   \nonumber
\]
Notice that both $g_{31}$ and $g_{32}$ are anti-PT-symmetric, and
$g_{33}$ is PT-symmetric.

The solvability conditions of Eq. (\ref{e:u3}) are
\[  \label{e:sol2}
\mbox{Im} \langle \phi_1, g_3 \rangle = \mbox{Re} \langle \phi_2, g_3 \rangle =0.
\]
Using the $g_3$ formula (\ref{e:g3}) and the symmetry properties of
the involved functions, these solvability conditions then yield a
condition for symmetry-breaking bifurcations as
\[  \label{e:cond1b}
\frac{\mbox{Im}\langle \phi_1, g_{31}\rangle}{\mbox{Im}\langle \phi_1, g_{32}\rangle}=\frac{\mbox{Re}\langle \phi_2, g_{31}\rangle}{\mbox{Re}\langle \phi_2, g_{32}\rangle}.
\]
Due to the PT symmetry of $\phi_1$ and anti-PT-symmetries of
$\phi_2,  g_{31}$ and $g_{32}$, $\langle \phi_1, g_{31}\rangle$ and
$\langle \phi_1, g_{32}\rangle$ are purely imaginary, and $\langle
\phi_2, g_{31}\rangle$, $\langle \phi_2, g_{32}\rangle$ are strictly
real. In addition, $\phi_2\ne i\phi_1$. Thus Eq. (\ref{e:cond1}) is
a non-trivial condition for symmetry-breaking bifurcations.

When we pursue this perturbation expansion to higher orders,
infinitely more non-trivial conditions will also appear (since these
calculations are straightforward, details are omitted here for
brevity). The fundamental reason for this infinite number of
conditions is that, due to the phase invariance of solitary waves,
when we solve the inhomogeneous $u_n$ equation, we can only
introduce one real parameter into the $u_n$ solution, which is the
coefficient of the $\psi$ term. But each $u_n$ equation has two
solvability conditions (since the kernel of $L_0^A$ has dimension
two), and neither solvability condition can be satisfied
automatically from symmetry considerations (for $n\ge 3$). This
means that we have twice as many solvability conditions as real
parameters. Because of this, we have an over-determined system for
real parameters, which results in an infinite number of non-trivial
conditions for symmetry-breaking bifurcations. In one spatial
dimension, Eq. (\ref{e:u}) does not admit any additional spatial
symmetries (except the PT symmetry). Due to this lack of additional
symmetries, it is practically impossible for these infinite
conditions to be satisfied simultaneously.

From the above analysis, we see that in a PT-symmetric potential,
the necessary condition for symmetry breaking, i.e.,
dim[ker($L_0$)]=2 with eigenfunction symmetries (\ref{e:sym1}), is
generically not satisfied. Even if that necessary condition is met,
symmetry breaking still requires infinitely more conditions to be
satisfied simultaneously, which is practically impossible for the 1D
system (\ref{e:u}). Thus, we conclude that symmetry breaking cannot
occur in the PT-symmetric system (\ref{e:u}).

Regarding the three specific examples
(\ref{e:realpot})-(\ref{e:Wx3}) in the PT system (\ref{e:up}) for
various values of $\epsilon$ (not necessarily small), we have found
numerically that the kernel of $L$ never contains a second
eigenfunction $[\psi, \psi^*]^T$ with anti-PT-symmetry at any $\mu$
value, thus the necessary condition for symmetry breaking is not
satisfied. This numerical finding corroborates our analytical result
that this necessary condition for symmetry breaking is generically
not met for a PT-symmetric potential.

\section{Summary and discussion}

In this article, we have investigated the possibility of continuous
families of non-PT-symmetric solitons in one-dimensional
PT-symmetric potentials. We have shown that families of asymmetric
solitons in a real symmetric potential are destroyed when this real
potential is perturbed by weak PT-symmetric perturbations. We have
also shown that in a general one-dimensional PT-symmetric potential,
symmetry breaking of PT-symmetric solitons cannot occur. This
contrasts real symmetric potentials where symmetry breaking of
solitary waves often takes place.

Based on these findings and Remark 1, we make the following
conjecture:

\vspace{0.07cm} \emph{The one-dimensional NLS equation (\ref{e:U})
with a complex PT-symmetric potential cannot admit continuous
families of non-PT-symmetric solitary waves.}

\vspace{0.1cm} Equivalently, this conjecture says that all
continuous families of solitary waves in a one-dimensional
PT-symmetric potential must be PT-symmetric.

The absence of continuous families of non-PT-symmetric solitons in
1D PT-symmetric potentials is an interesting phenomenon, since it
contrasts real symmetric potentials, where families of asymmetric
solitons often exist. This means that, even though PT-symmetric
potentials can support continuous families of solitons, which makes
such dissipative potentials analogous to conservative real
potentials, the types of soliton families allowed by PT-symmetric
potentials are nonetheless limited. So the dissipative nature of a
PT-symmetric potential does leave its signature on the structure of
its solitary waves, and this signature distinguishes PT-symmetric
potentials from real symmetric ones.

We would like to point out that the above conjecture does not
exclude the possibility of 1D PT-symmetric potentials supporting
\emph{isolated} non-PT-symmetric solitons (i.e., non-PT-symmetric
solitons existing at isolated propagation-constant values). In a
certain finite-dimensional PT-symmetric system (the quadrimer
model), isolated non-PT-symmetric solutions have been reported (Li
\& Kevrekidis 2011). In the 1D NLS equation (\ref{e:U}) with a
PT-symmetric potential, such isolated non-PT-symmetric solitons can
also exist, as our preliminary numerics has shown. These isolated
solitons are reminiscent of dissipative solitons in the
Ginzburg-–Landau and other dissipative equations (Akhmediev \&
Ankiewicz 2005), and they can coexist with continuous families of
solitons in a PT-symmetric potential.

The analytical results in this article can be extended to a large
class of higher-dimensional NLS equations with PT-symmetric
potentials, but not to all of them. In higher spatial dimensions,
the PT symmetry of a potential is compatible with certain other
spatial symmetries, such as $x$-symmetry or $y$-symmetry. For
instance, we can easily construct two-dimensional complex potentials
$V(x,y)$ with the following PT-symmetry as well as $x$-symmetry,
\[
V^*(x,y)=V(-x,-y), \quad V(x,y)=V(-x, y),  \nonumber
\]
i.e., the real part of the potential is symmetric in both $x$ and
$y$, but the imaginary part of the potential is symmetric in $x$ and
anti-symmetric in $y$. Due to this additional $x$-symmetry,
continuous families of non-PT-symmetric solitons \emph{can} exist in
this 2D PT-symmetric potential, and symmetry breaking of
PT-symmetric solitons \emph{can} occur. Putting this 2D problem in
the framework of the earlier analysis in this article, the reason
for the existence of PT-symmetry breaking and families of
non-PT-symmetric solitons in this 2D PT potential is that, due to
the additional $x$-symmetry of the potential and its ramifications
for the symmetries of the underlying solitons and eigenfunctions in
the kernels of the linearization operators, those infinite
conditions in our earlier analysis can now be all satisfied. Details
on this 2D problem will be reported elsewhere. However, if this 2D
PT-symmetric potential $V(x,y)$ does not admit those additional
spatial symmetries (such as $x$-symmetry and $y$-symmetry), then the
analysis in this article would still apply, and families of
non-PT-symmetric solitons still cannot be expected. Thus, absence of
continuous families of non-PT-symmetric solitons in PT-symmetric
potentials is not restricted to one spatial dimension, but holds for
most higher-dimensional PT-symmetric potentials as well.

\vspace{0.2cm} \textbf{Acknowledgment:} The author thanks Prof.
Vladimir Konotop for very helpful discussions. This work was
supported in part by the Air Force Office of Scientific Research
(Grant USAF 9550-12-1-0244) and the National Science Foundation
(Grant DMS-1311730).

\bigskip

\catcode`\@ 11
\def\journal#1&#2,#3 {\begingroup \let\journal=\d@mmyjournal {\frenchspacing\it #1\/\unskip\,}
{\bf\ignorespaces #2}\rm, #3\endgroup}
\def\d@mmyjournal{\errmessage{Reference foul up: nested \journal macros}}

\end{document}